\documentclass[12pt]{article}
\usepackage{amsmath}
\usepackage{graphicx}

\begin{document}

\begin{center}

{\bf\large{A Brief History of Mass}}\\

\bigskip

Scott Willenbrock \\

\bigskip

Department of Physics, University of Illinois at Urbana-Champaign \\
1110 West Green Street, Urbana, Illinois 61801, USA \\

\bigskip

willen@illinois.edu

\end{center}

\medskip

\noindent {\bf Abstract}: It has been known since the 1950s that an unstable particle is associated with a complex pole in the propagator. This had to be rediscovered twice: in the early 1970s in the context of hadronic resonances, and in the early 1990s in the context of the $Z$ boson. The physical mass of the particle is the real part of the pole in the complex energy plane. In hadronic physics, this replaced the ``Breit-Wigner mass,'' which was found to depend on the parameterization of the ``energy-dependent width.'' In $Z$ physics, it replaced the ``on-shell'' mass, which was found to be gauge dependent. Although the mass defined from the complex pole position has been widely discussed in the literature, it has not yet made its way into quantum field theory textbooks.

\bigskip\medskip

\noindent {\bf 1. Introduction}

\bigskip

I recently realized \cite{W1} that the mass (and decay rate) of an unstable particle has a simple, unique, and unambiguous definition, and that this definition has been known for over sixty years. This definition is widely known and used in the hadronic physics community, but it appears to be little known in the electroweak physics community, with the exception of neutral meson oscillations. How did this strange situation come to pass?

This article is an attempt to trace the history of the mass of an unstable particle.
I set the stage with a discussion of our modern understanding of unstable particles. I then go back and review the early theoretical treatment of unstable particles and its application to hadronic resonances. Next I leap forward two decades to the $Z$ resonance, and argue that the history of the theoretical treatment of the $Z$ is a repetition of what happened with hadronic resonances. I then conduct a quick review of the treatment of unstable particles in several quantum field theory textbooks.

Particle physics, like any other scientific field, progresses clumsily, with years of confusion followed by moments of clarity. Vestiges of the confusion sometime remain. It is healthy to continuously ask if we are describing our current understanding in the best possible way. While we have a good understanding of unstable particles, I believe this knowledge is not adequately disseminated.

\bigskip

\noindent {\bf 2. Modern Theory}

\bigskip

We begin with a discussion of the modern way to treat unstable particles in quantum field theory. I will mostly refrain from citing any literature in this section, and include the relevant citations in later sections. For further details, see Ref.~\cite{W1}.

Let's begin by considering the propagator of a complex scalar field; a similar calculation can be carried out for any field. In a free field theory, the propagator is the amplitude for a particle to be created at the origin and destroyed at spacetime point $x$ (or an antiparticle to be created at $x$ and destroyed at the origin),
\begin{equation}
\langle 0 | T \phi(x) \phi^\dagger (0) |0\rangle = \int \frac{d^4p}{(2\pi)^4}\frac{ie^{-ip\cdot x}}{p^2-m^2+i\epsilon}\;.
\end{equation}
We now include interactions to all orders in perturbation theory.
Consider the self-energy corrections to the momentum-space propagator from the diagrams in Fig.~1, where the shaded circle denotes the bare one-particle-irreducible self-energy to all orders in perturbation theory, $i\Pi(p^2)$.
Summing the series of diagrams gives
\begin{eqnarray}
&&\frac{i}{p^2 - m_B^2+i\epsilon}\left(1-\frac{\Pi(p^2)}{p^2-m_B^2+i\epsilon}+\cdots\right)\\
&&=\frac{i}{p^2 - m_B^2 + \Pi(p^2)+i\epsilon}
\label{bare}
\end{eqnarray}
where $m_B$ is the bare mass.
The pole of the full momentum-space propagator, $\mu^2$, is the solution to the equation
\begin{equation}
\mu^2-m_B^2+\Pi(\mu^2)=0\;.
\label{pole}
\end{equation}
Combining Eqs.~(\ref{bare}) and (\ref{pole}) yields
\begin{equation}
\langle 0 | T \phi(x) \phi^\dagger (0) |0\rangle = \int \frac{d^4p}{(2\pi)^4}\frac{ie^{-ip\cdot x}}{p^2-\mu^2+\Pi(p^2)-\Pi(\mu^2)+i\epsilon}
\label{bare2}
\end{equation}
If the particle is unstable $\Pi(\mu^2)$ is complex, and hence, from Eq.~(\ref{pole}), $\mu^2$ is complex. The pole position is gauge invariant and infrared safe.

The scattering amplitude of a process with an intermediate unstable particle acquires a complex pole from the propagator,
\begin{equation}
S \sim \frac{1}{p^2-\mu^2}\;.
\label{LOprop}
\end{equation}
Going to the rest frame of the unstable particle we can rewrite Eq.~(\ref{LOprop}) as
\begin{equation}
S \sim \frac{1}{p_0^2-\mu^2}=\frac{1}{(p_0 - \mu)(p_0 + \mu)}\;.
\label{Epole}
\end{equation}
To find the time dependence of the scattering amplitude, we Fourier transform from energy to time:
\begin{equation}
S \sim \int_{-\infty}^\infty dp_0\;  \frac{e^{-ip_0t}}{(p_0 - \mu)(p_0 + \mu)}\;.
\label{FT}
\end{equation}
For $t>0$ we can close the contour in the lower half complex $p_0$ plane, as shown in Fig.~2, since the integral along the large semicircle is exponentially damped. Using the residue theorem, we pick up the contribution of the pole at $p_0 = \mu$,
\begin{equation}
S \sim e^{-i\mu t} = e^{-i{\rm Re}\,\mu t}e^{{\rm Im}\,\mu t}\;.
\label{time}
\end{equation}
The decay probability is given by the square of the scattering amplitude,
\begin{equation}
|S|^2 \sim e^{2{\rm Im}\,\mu t}\;.
\end{equation}
This corresponds to exponential decay with $2{\rm Im}\,\mu = -\Gamma$, where $\Gamma$ is the decay rate. Hence we learn that ${\rm Im}\,\mu = - \Gamma/2$.

For $t<0$ we close the contour in the upper half complex $p_0$ plane and pick up the contribution of the pole at $p_0 = -\mu$. This pole is associated with antiparticle propagation. The residue theorem gives $S \sim e^{i{\rm Re}\,\mu t}e^{-{\rm Im}\,\mu t}$, and hence $|S|^2 \sim e^{\Gamma t}$. This also corresponds to exponential decay since $t$ is negative, and proves that the particle and antiparticle have the same lifetime.

The oscillatory frequency of the unstable particle is dictated by the particle energy, which in the rest frame is just its mass.  Hence ${\rm Re}\,\mu = m$, and we conclude that
\begin{equation}
\boxed{\mu = m - \frac{i}{2}\Gamma}\;.
\label{mGamma}
\end{equation}
Eq.~(\ref{mGamma}) provides an unambiguous decomposition of $\mu$ into physically meaningful quantities. For this reason we will refer to $m$ as the physical mass.

Returning to Eq.~(\ref{pole}), and using Eq.~(\ref{mGamma}), we find
\begin{equation}
{\rm Im}\,\mu^2 = -m\Gamma = - {\rm Im}\, \Pi(\mu^2)
\end{equation}
or
\begin{equation}
\Gamma = \frac{1}{m}{\rm Im}\, \Pi(\mu^2)\;.
\label{width}
\end{equation}
This is an implicit formula for the decay rate, since $\mu = m - \frac{i}{2}\Gamma$, and can be solved by expanding in powers of $\Gamma/m$,
\begin{equation}
m\Gamma  = Z\,{\rm Im}\,\Pi(m^2)\left(1-\frac{1}{2}{\rm Im}\,\Pi(m^2){\rm Im}\,\Pi^{\prime\prime}(m^2)
-\frac{1}{4m^2}{\rm Im}\,\Pi(m^2){\rm Im}\,\Pi^\prime(m^2)+\cdots\right)
\label{mG}
\end{equation}
where $Z^{-1} = 1+{\rm Re}\,\Pi^\prime(m^2)$ is the analogue of the field renormalization constant that would be necessary if the particle were an asymptotic state.

The leading term in Eq.~(\ref{mG}), $m\Gamma={\rm Im}\,\Pi(m^2)$, is the familiar leading-order expression. That expression is usually derived by treating the decaying particle as an asymptotic state, which is clearly an approximation valid only for $\Gamma \ll m$. The exact expression, without any approximations, is given by Eq.~(\ref{width}).

Another feature of the propagator is a branch cut along the positive real axis in the complex energy plane, with a branch point at the threshold energy for the decay. This is because the function $\Pi(p^2)$ has such a branch cut, where the nature of the branch point is $q^{2L+1}$, where $q$ is the momentum of each decay product (for a two-body decay) and $L$ is their angular momentum. The function $\Pi(p^2)$ is double-valued, and is best thought of as having two sheets (called Riemann sheets) connected at the branch cut, the first sheet corresponding to ${\rm Im}\,q>0$ and the second sheet corresponding to ${\rm Im}\,q<0$.  The pole in the propagator is on the second sheet. The branch cut is associated with the nonresonant part of the scattering amplitude, and does not affect the exponential decay of the unstable particle.

In the energy region near the pole at $p_0 = \mu$, the amplitude of Eq.~(\ref{Epole}) can be approximated by neglecting the antiparticle pole at $p_0 = -\mu$,
\begin{equation}
S \sim \frac{1}{E - \mu}
\end{equation}
where $E=p_0$.
The cross section in the resonance region is thus approximately
\begin{equation}
|S|^2 \sim \frac{1}{(E-m)^2 + \Gamma^2/4}
\label{BW}
\end{equation}
which is the well-known Breit-Wigner formula \cite{BW}. The resonance shape has a full width at half maximum of $\Gamma$, which is why the decay rate $\Gamma$ is also called the width. A similar formula may also be obtained from nonrelativistic quantum mechanics, such as in the original paper. For this reason Eq.~(\ref{BW}) is sometimes referred to as a ``nonrelativistic Breit-Wigner,'' but we avoid this terminology since the equation is also true at relativistic energies.

Another way to approximate the cross section in the resonance region is to start from Eq.~(\ref{LOprop}). Making the approximation, valid for $\Gamma \ll m$,
\begin{equation}
\mu^2 = \left(m-\frac{i}{2}\Gamma\right)^2 \approx m^2-im\Gamma\;\;\;\;\;\;(\Gamma\ll m)
\label{mGammaapprox}\end{equation}
gives the propagator
\begin{equation}
S \sim \frac{1}{p^2 - m^2 + im\Gamma}\;\;\;\;\;\;(\Gamma\ll m)
\label{approxprop}\end{equation}
and hence the cross section
\begin{equation}
|S|^2 \sim \frac{1}{(p^2 - m^2)^2 + m^2\Gamma^2}\;.\;\;\;\;\;\;(\Gamma\ll m)
\label{RIBW}
\end{equation}
This is often referred to as a relativistic Breit-Wigner formula. The manifest Lorentz invariance of this expression follows from the inclusion of both particle and antiparticle poles.

The formula $\mu^2 = m^2 - im\Gamma$ [see Eq.~(\ref{mGammaapprox})], which has gained widespread currency, is also an approximation. {\em The precise definition of the mass and width of an unstable particle is $\mu = m - (i/2)\Gamma$ [Eq.~(\ref{mGamma})].}

Inserting Eq.~(\ref{mGamma}) into Eq.~(\ref{LOprop}) without making any approximations gives the propagator
\begin{equation}
S \sim \frac{1}{p^2 - \mu^2} = \frac{1}{p^2 - \left(m - \frac{i}{2}\Gamma\right)^2}
\label{propagator}
\end{equation}
and hence the cross section
\begin{equation}
\boxed{|S|^2 \sim \frac{1}{(p^2 - m^2 + \Gamma^2/4)^2+m^2\Gamma^2}}\;.
\label{resonance}
\end{equation}
It may seem unconventional to include the $\Gamma^2/4$ term in the denominator of the above expression, but its presence follows from the precise definition of mass and width provided by Eq.~(\ref{mGamma}). Eq.~(\ref{resonance}) is a relativistic Breit-Wigner formula, but without making the approximation $\Gamma \ll m$ as in Eq.~(\ref{RIBW}). Including the $\Gamma^2/4$ term is numerically significant; for example, for the $\rho$ meson it increases the mass by 3.5 MeV, and for the $Z$ boson it increases the mass by 8.6 MeV, both of which are greater than the experimental uncertainty. However, for most particles (including these two), Eq.~(\ref{RIBW}) is generalized by replacing $\Gamma$ with an ``energy-dependent width,'' as we shall discuss later. This generally results in the physical mass being less than the mass in Eq.~(\ref{RIBW}), rather than greater.

\bigskip\

\noindent {\bf 3. Early Theory}

\bigskip

It was Peierls who first proposed, in 1954, that an unstable particle be associated with a complex pole on the second Riemann sheet of the propagator \cite{P}. He studied a one-dimensional, nonrelativistic quantum-mechanical model to support his conjecture. This proposal was pursued several years later by L\'evy \cite{L} in the context of the Lee model \cite{Lee}, which is a nonrelativistic quantum field theory that is exactly solvable in sectors with just one or two particles. He was building on the work of Glaser and K\"allen \cite{GK}, who had begun the study of unstable particles in the Lee model. L\'evy concluded that the Lee model supports Peierls' conjecture. The Lee model was chosen for these studies because it yielded results that do not depend on perturbation theory.

Around the same time, Matthews and Salam \cite{MS} proposed defining the mass of an unstable particle in terms of an average of the spectral function of K\"allen \cite{Kallen} and Lehmann \cite{Lehmann}. Based on the discussion after a talk by Salam at ICHEP 58 \cite{Salam}, it is clear that neither Peierls nor Gell-Mann looked very favorably upon this definition. Both spoke in favor of using the pole position to define the mass of an unstable particle.

It was not until 1961 that Jacob and Sachs \cite{JS} studied unstable particles using perturbation theory. Using a simple one-loop calculation, they found that the mass (and decay rate) of an unstable particle does indeed correspond to the pole of the propagator on the second sheet of the energy-squared plane,\footnote{One of the dangers of working in the energy-squared plane is that it leads to typos. The authors conclude that the mass of the particle is given by ${\rm Re}\, z_0$, and the lifetime by $(2{\rm Im}\, z_0)^{-1}$, where $z_0$ is the pole in the energy-squared plane. Neither of these conclusions make sense dimensionally. It is clear from the discussion in the article that they meant $\sqrt{z_0}$ in place of $z_0$ in these expressions (actually, they meant $-\sqrt{z_0}$ in the case of the lifetime). The formula $z_0 = (m - \frac{i}{2}\Gamma)^2$ is explicitly stated a few years later in a paper by Harte and Sachs \cite{HS}, see their Eq.~(26).} $\mu^2 = (m - \frac{i}{2}\Gamma)^2$. Unlike L\'evy, they did not make the approximation that the decay rate is much less than the mass.

Let us pause and ask which unstable particles these authors had in mind when writing these papers. In the 1950s, almost all of the known unstable particles decayed via the electroweak interactions, and had widths many orders of magnitude less than their mass. The one exception was the so-called (3,3) resonance, which decayed via the strong interaction and had a width of about 10$\%$ of its mass; we will discuss this particle further in the next section. Given that the authors were mostly interested in describing the electroweakly-decaying particles, it is surprising that it was not until 1961 that anyone thought to apply perturbation theory to the problem. It reflects the distrust in perturbative quantum field theory, applied to the weak and strong interactions, that was present at the time. The popular two-volume textbook by Bjorken and Drell \cite{BD} from the mid 1960s concludes with these ominous words: ``... conclusions based on the renormalization group arguments concerning the behavior of the theory summed to all orders are dangerous and must be viewed with due caution. {\em So is it with all conclusions from local relativistic field theories}''(my emphasis).

After the 1961 paper of Jacob and Sachs, everything was in place for the particle physics community to recognize that the mass and decay rate of an unstable particle are obtained from the pole in the particle's propagator, $\mu^2 =  (m - \frac{i}{2}\Gamma)^2$. But that is not how things unfolded. That same year witnessed the discovery of the $\rho$ meson \cite{EMWW}, one of the first strongly-decaying mesons, at the Brookhaven Cosmotron. This prompted Gell-Mann and Zachariasen \cite {GZ} to consider the treatment of unstable particles in quantum field theory. They used an exactly-solvable model field theory of Zachariasen \cite{Z}, similar to the Lee model, to explore the topic. In this model, the $\pi\pi$ scattering amplitude has a denominator that is similar to that of a propagator, and they defined the squared mass of the $\rho$ meson to be the $p^2$ at which the real part of the denominator vanishes. They then paused to show that, {\it for small widths} ($\Gamma \ll m$), this squared mass is equal to the real part of the pole in the complex $p^2$ plane, and that the pole position is given by $\mu^2 = m^2 - im\Gamma$ ({\it again, assuming} $\Gamma \ll m$). They explicitly stated that, for larger widths, their definition of mass and width would not be the same as the mass and width defined in terms of the pole on the second sheet. They then ``abandon the second sheet'' and continue to use their definition of mass in terms of the zero of the real part of the denominator. The denominator $p^2-m^2+im\Gamma+\cdots$ appears in several expressions later in the paper, which explicitly uses their definition of mass and implicitly assumes $\Gamma \ll m$.

From a historical perspective, this paper had two consequences. First, it gave legitimacy to defining the mass of an unstable particle as the zero of the real part of the denominator of the propagator. Second, it introduced the approximation $\mu^2 = (m-\frac{i}{2}\Gamma)^2 \approx m^2 -im\Gamma$, even though the $\rho$ meson is not a particularly narrow resonance ($\Gamma/m \sim 0.2$). Their dismissive attitude towards the second sheet may also have had an impact.

Let's go back and reconsider the propagator of Eq.~(\ref{bare}). Following Gell-Mann and Zachariasen, let's define the squared mass as the zero of the real part of the denominator of the propagator,
\begin{equation}
M^2 - m_B^2 + {\rm Re}\,\Pi(M^2) = 0
\label{mBW}
\end{equation}
Using this to eliminate the bare mass yields the propagator
\begin{equation}
\frac{1}{p^2-M^2 + \Pi(p^2) - {\rm Re}\,\Pi(M^2)}\approx \frac{1}{p^2-M^2 + i{\rm Im}\,\Pi(p^2)}
\label{BJpropagator}
\end{equation}
where, in the final expression, we have neglected ${\rm Re}\,\Pi(p^2) - {\rm Re}\,\Pi(M^2)$ in the resonance region. We know that the function $\Pi(p^2)$ has a branch point on the real axis at the threshold for the particle decay, and that the nature of the branch point is $q^{2L+1}$, where $q$ is the momentum of each decay product (for a two-body decay) and $L$ is their angular momentum. Other than that, the function ${\rm Im}\,\Pi(p^2)$ is model dependent. The derivation above was presented in a 1963 paper by Bordes and Jouvet \cite{BJ}.

The particle's decay rate [in the approximation $\Gamma\ll m$, see Eq.~(\ref{mG})] is given by $m\Gamma = {\rm Im}\;\Pi(m^2)$. This suggests the notation\footnote{Alternatively, ${\rm Im}\;\Pi(p^2)=\sqrt{p^2}{\it\Gamma} (p^2)$}  ${\rm Im}\;\Pi(p^2)=M{\it\Gamma} (p^2)$, where ${\it\Gamma}(p^2)$ is called the ``energy-dependent width.''  Thus the propagator is written
\begin{equation}
\frac{1}{p^2-M^2 + iM{\it\Gamma}(p^2)}
\label{Propagator}
\end{equation}
which, for constant ${\it\Gamma}$, is identical to the expression written by Gell-Mann and Zachariasen. The square of this propagator
\begin{equation}
\frac{1}{(p^2-M^2)^2 + M^2{\it\Gamma}^2(p^2)}
\label{RBW}
\end{equation}
came to be known as a ``relativistic Breit-Wigner,'' and the mass as the ``Breit-Wigner mass.'' This equation can be found in a 1964 paper by Dave Jackson \cite{Jackson} entitled ``Remarks on the Phenomenological Analysis of Resonances,'' which references Bordes and Jouvet,  but I suspect that Eq.~(\ref{RBW}) was already in the air at the time. For example, essentially the same formula is given at the end of the 1958 talk of Salam mentioned above \cite{Salam}. Jackson presents no derivation of this equation, but makes the cryptic remarks that Eq.~(\ref{RBW}) ``cannot quite be derived by perturbation theory involving a simple propagator for the resonance. The resonance must be given a complex mass or the propagator must be calculated more accurately \cite{BJ}. But justification for the denominator [of Eq.~(\ref{RBW})] is hardly necessary.'' I believe that this paper, with over 700 citations, cemented the relativistic Breit-Wigner in physicists' minds.\footnote{Jackson wrote this paper while he was on leave (at CERN) from my own institution.} The paper of Jacob and Sachs \cite{JS}, in contrast, has less than 80 citations.

As evidence of how ingrained the relativistic Breit-Wigner became, consider the 1967 book of Pilkuhn on {\sl The Interactions of Hadrons} \cite{Pilkuhn}. In Chapter 2, Eq.~(3.13), he shows that the relativistic propagator is equal to the sum of a Breit-Wigner for the particle and the antiparticle as in Eq.~(\ref{Epole}), but in order to obtain the propagator of Eq.~(\ref{approxprop}) (or Eq.~(\ref{Propagator}) for constant ${\it\Gamma}$) he makes the approximation $\Gamma \ll m$, even though the book is concerned with strongly-decaying hadrons. If he had not made that approximation he would have been led to the exact result, Eq.~(\ref{propagator}). Thus it seems that the approximation $\Gamma \ll m$ was implicitly and explicitly part of physicists' mindset, even when dealing with hadronic resonances.

\bigskip

\noindent {\bf 4. The (3,3) Resonance}

\bigskip

In 1952, Fermi's group studied pion-nucleon scattering at the Chicago cyclotron and observed cross sections that rise rapidly with increasing energy \cite{AFLN}. Brueckner interpreted these results as a resonance in the $J=3/2$, $I=3/2$ channel that came to be known as the (3/2,3/2) [or (3,3)] resonance \cite{B}. The nature of this resonance was unclear. It would be two decades before it became commonly accepted that it is a bound state of three quarks, the particle we now call the $\Delta (1232)$ baryon. It would prove to be the first of the many strongly-decaying particles discovered in the 1960s and beyond.\footnote{It is interesting to trace the attitude of the particle physics community towards the $\Delta(1232)$ baryon, as well as other particles, by way of its treatment in the Review of Particle Physics, published by the Particle Data Group \cite{RPP}. This particle is not listed in the first incarnation of this review in 1958 by Barkas and Rosenfeld, while the sixteen listed particles are referred to as ``elementary particles.'' The only mention of the (3/2,3/2) resonance is in a table on Atomic and Nuclear Constants. The (3/2,3/2) resonance is first listed in the 1961 edition, along with a few other strongly-decaying hadrons, in a table of ``Possible Resonances of Strongly Interacting Particles,'' and is given the name $N^*(1238)$. This list is greatly expanded in the 1963 edition in a table of ``Tentative Data on Strongly Interacting Particles and Resonances,'' which includes all known hadrons, whether stable, electroweakly decaying, or strongly decaying. The sixteen ``elementary particles'' listed in the 1958 review have been downgraded to simply ``particles.'' In the 1964 edition that list has grown to nineteen particles, which are now called ``stable particles,'' by which is meant particles that do not decay via the strong interaction. The hadron list has also grown, and is simply called ``mesons'' and ``baryons.'' In the text, the hadrons that decay strongly are referred to as ``meson and baryon resonances'' and, for the first time, as ``unstable particles.'' This edition is called ``Data on Elementary Particles and Resonant States,'' and for the first time the (3/2,3/2) resonance is referred to by its modern name, $\Delta(1238)$. The ``Breit-Wigner mass'' was lowered to 1236 MeV in 1965 and to 1232 MeV in 1974, where it remains to this day.}

In Brueckner's paper he described the (3,3) resonance using the Breit-Wigner formula, Eq.~(\ref{BW}), which has two parameters, $m$ and $\Gamma$. As the data became more precise this formula was no longer sufficient, so several three-parameter extensions of the Breit-Wigner formula became popular, one being the relativistic Breit-Wigner [Eq.~(\ref{RBW})] with a particular model for ${\it\Gamma}(p^2)$, and two being Breit-Wigner functions with particular models for ${\it\Gamma}(E)$ (the Breit-Wigner formula having been generalized to include an ``energy-dependent width,'' just like its relativistic counterpart). The mass and width extracted from these different models were in close agreement. However, later it became necessary to add a nonresonant background term in order to fit the data, and it was found that the mass and width extracted from these different models no longer agreed \cite{Barbaro}.

As recounted by Barbaro-Galtieri \cite{Barbaro}, it was Sidney Coleman who suggested at the 1971 Erice summer school that the (complex) pole position of these various models should all agree. This was confirmed by Ball, Campbell, Lee, and Shaw \cite{BCLS}. Thus, for the first time (since Brueckner), the mass and width of a hadronic resonance were defined in terms of the pole in the complex energy plane, $\mu = m - \frac{i}{2}\Gamma$. Coleman was well aware of the fundamental nature of the pole position since he had lectured on this topic at the same summer school three years earlier \cite{ColemanErice}. The mass of the $\Delta(1236)$, defined in terms of the pole position, was first reported in the 1972 Review of Particle Properties \cite{RPP} as $M_{pole} = 1211.11$ MeV. The $\Delta(1236)$ continued to derive its name from the mass parameter in a relativistic Breit-Wigner fit, and two years later it was renamed the $\Delta(1232)$, which it retains to this day.

Once introduced, the idea of using the pole position to define the mass and width of an unstable particle spread. It got a big boost from the 1975 paper of Longacre {\it et al.} \cite{LRLSCL}, which led to many $N$ and $\Delta$ resonances having their pole positions reported in the 1976 Review of Particle Properties \cite{RPP}. However, it wasn't until 40 years later (2016) that the pole position was informally recognized as {\it the} fundamental property of an unstable particle by listing it first, ahead of the Breit-Wigner mass and width, in the particle listings. The fundamental nature of the pole position had been emphasized a few years earlier in the very first review on ``Resonances,'' in the 2013 Review of Particle Physics.

A similar story played out in the meson sector. The first meson whose pole position was listed (in the 1973 Review of Particle Properties \cite{RPP}) was the $S^*(1000)$ meson [today called the $f_0(980)$], for the reason that the $I=0$, $S$-wave $\pi\pi$ scattering amplitude was ``much too complicated to allow a description in terms of one or several Breit-Wigner resonances.'' However, it wasn't until nearly 50 years later (2021) that the pole position of mesons was listed ahead of the Breit-Wigner masses and widths. In the case of the $\rho$ meson, the model independence of the pole position was demonstrated in 1999 by Benayoun, O'Connell, and Williams \cite{BOW}.

\bigskip

\noindent {\bf 5. The $Z$ Boson - History Repeats Itself}

\bigskip

In the late 1970s, planning began for the Large Electron-Positron collider (LEP) at CERN and the Stanford Linear Collider (SLC), both $e^+e^-$ colliders operating near the $Z$ boson mass (and above in the case of LEP). These would be the first machines to resolve a resonance that could be calculated perturbatively, unlike the hadronic resonances that had come before. The theoretical community was faced with the challenge of calculating the resonant cross section with enough precision to match that of the experiment.

Physicists had already established methods to calculate electroweak corrections to low-energy processes such as muon decay, $\beta$ decay, neutrino-induced processes, {\it etc}. An early renormalization scheme was proposed by Sirlin \cite{Sirlin}, and came to be known as the ``on-shell'' scheme because it relied on physical processes to define the renormalized couplings.  In particular, the weak mixing angle was defined by $\sin^2\theta_W = 1-\frac{M_W^2}{M_Z^2}$, where $M_W$ and $M_Z$ are the $W$ and $Z$ boson masses. The ``on-shell'' scheme is still widely used, as evidenced by its treatment in the review ``Electroweak Model and Constraints on New Physics'' in the 2024 Review of Particle Physics \cite{RPP2024}.

Sirlin defined the $W$ and $Z$ boson masses via
\begin{equation}
M^2 - m_B^2 + {\rm Re}\,\Pi(M^2) = 0\;.
\label{OS}
\end{equation}
This is identical to the definition of the squared mass of an unstable particle as the zero of the real part of the denominator of the propagator, Eq.~(\ref{mBW}), called the ``Breit-Wigner mass.'' History repeats itself.

Let's evaluate the difference between the ``on-shell'' mass and the physical mass obtained from the complex pole in the propagator, Eq.~(\ref{pole}):
\begin{eqnarray}
M^2 - m^2 & = & {\rm Re}\,\Pi(\mu^2) - {\rm Re}\,\Pi(M^2) - \frac{1}{4}\Gamma^2 \\
& = & m\Gamma\,{\rm Im}\,\Pi^\prime(m^2) - \frac{1}{4}\Gamma^2 + \cdots
\label{dif}
\end{eqnarray}
where we have performed a Taylor expansion of $\Pi(\mu^2)$ about $m^2$ in the final expression. We see that the difference between the two masses is of next-to-next-to-leading order (NNLO) in $\Gamma/m$, so for low-energy processes calculated at next-to-leading-order (NLO) the difference is inconsequential, although it is relevant at NNLO \cite{FHWW}.

The ``on-shell' mass was adopted by the groups calculating the radiative corrections to the $Z$ boson line shape. As in Eq.~(\ref{Propagator}), this led to a propagator with an ``energy-dependent width'',
\begin{equation}
\frac{1}{p^2-M^2 + ip^2{\it\Gamma}/M}
\label{Edepwidth}
\end{equation}
since ${\rm Im}\,\Pi(p^2) \sim p^2$ for a $Z$ boson coupled to light fermions. An exception to this was a contribution to a CERN Yellow Report in 1986 by Consoli and Sirlin \cite{CS}, who defined the $Z$ boson mass via the complex pole in the propagator, $\mu^2 = m_Z^2 - im_Z\Gamma_Z$.  The fact that they used this approximate expression, rather than the exact expression $\mu = m_Z-\frac{i}{2}\Gamma_Z$, demonstrates again how ingrained it had become. They referred to this approximate expression as ``customary.''

The first indication that there is something amiss with the ``on-shell'' mass came from Valencia and Willenbrock \cite{VW}, who showed that the ``on-shell'' mass of a (hypothetical) heavy Higgs boson depends on the choice of fields. In contrast, they found that the pole position is field-redefinition invariant, and argued that it must also be gauge invariant \cite{WV}. They suggested that the $Z$ boson mass be defined by the exact expression $\mu = m_Z-\frac{i}{2}\Gamma_Z$. Around the same time, Stuart \cite{Stuart} also realized that the pole position provides a gauge-invariant definition of the $Z$ boson mass. He proposed the approximate expression $\mu^2 = m_Z^2 - im_Z\Gamma_Z$, which he referred to as ``conventional.'' He later revisited the question of the definition of the $Z$ boson mass via the exact expression versus the approximate expression and concluded that ``There is no good physical reason to prefer one over the other...'' \cite{Stuart2}, because he did not consider the time dependence of the amplitude.\footnote{One could choose to define the $Z$ mass and width via $\mu^2 = m_Z^2 - im_Z\Gamma_Z$, but then they lose their physical meaning. Furthermore, all hadron masses are defined via $\mu = m - \frac{i}{2}\Gamma$.}

Shortly after these papers appeared, Sirlin \cite{Sirlin2} realized that the gauge-invariance of the pole position implies that the ``on-shell'' mass is gauge dependent and therefore unphysical. One can see this by inspecting the difference between the mass defined by the pole position and the ``on-shell' mass, Eq.~(\ref{dif}). The term proportional to ${\rm Im}\,\Pi^{\prime}(m^2)$ is gauge dependent (in $R_\xi$ gauge) due to loops of charged Goldstone bosons, which contribute at one loop if their mass is less than half the $Z$ boson mass. Thus the ``on-shell'' mass is gauge dependent at NNLO. In the class of gauges where the charged Goldstone bosons are heavier than half the $Z$ boson mass, the gauge dependence appears at NNNLO.

It was too late for this work to have any impact on the LEP and SLC experiments. The experiments fit the resonance data to a cross section with an ``energy-dependent width'' [Eq.~(\ref{Edepwidth})], which is the standard to this day (see the review on the ``$Z$ Boson'' in the 2024 Review of Particle Physics \cite{RPP2024}). Fortunately, the resonant amplitude, Eq.~(\ref{Edepwidth}), still has a simple pole, so one can solve for the pole position and relate it to the physical $Z$ boson mass. One finds
\begin{equation}
m = M \left(\frac{r+1}{2r^2}\right)^{\frac{1}{2}}\approx M\left(1-\frac{3}{8}\left(\frac{{\it \Gamma}}{M}\right)^2\right)
\label{m}
\end{equation}
where $r=\left(1+\left({\it \Gamma}/M\right)^2\right)^{\frac{1}{2}}$.
Using the values $M_Z = 91.1880 \pm 0.0020\; {\rm GeV}$ and ${\it\Gamma}_Z = 2.4955 \pm 0.0023\; {\rm GeV}$ from the 2024 Review of Particle Physics \cite{RPP2024} yields the physical mass \begin{equation}
m_Z = 91.1624 \pm 0.0020\; {\rm GeV}
\end{equation}
which is 25.6 MeV less than $M_Z$.

The physical mass of the $Z$ boson was reported in the very first ``Note on the $Z$ Mass'' in the 1992 Review of Particle Properties \cite{RPP}. A much more extensive ``Note on the $Z$ Boson'' was written for the 1994 Review, and included an ``$S$-matrix approach to the $Z$.'' In this approach, the $Z$ mass and width are defined in terms of the pole position using the approximate formula $\mu^2 = m_Z^2 - im_Z\Gamma_Z$ \cite{LRR}, again demonstrating how ingrained this formula had become. The exact definition, $\mu = m_Z - \frac{i}{2}\Gamma$, is also mentioned. This remained the situation for many years.

In 2005, Bohm and Sato \cite{BS} argued that it is problematic to have three different definitions of the $Z$ boson mass. They defined state vectors for unstable particles in quantum field theory, and were led to the exact result, $\mu = m - \frac{i}{2}\Gamma$. A few years later, in 2008, the Review of Particle Physics \cite{RPP} dropped this definition from the review on the ``$Z$ Boson'' without comment, and this remains the situation today. Thus there are now two definitions of the $Z$ mass in the Review, neither of which are the physical one.

\bigskip

\noindent {\bf 6. The $W$ Boson, Top Quark, and Higgs Boson}

\bigskip

The world-average\footnote{There is a recent $W$ mass measurement from CMS \cite{CMSWmass} that is not yet included in the world average.} values $M_W = 80.3692 \pm 0.0133\; {\rm GeV}$ and ${\it \Gamma}_W = 2.085 \pm 0.042\; {\rm GeV}$ from the 2024 Review of Particle Physics \cite{RPP2024} yield the physical $W$ boson mass via Eq.~(\ref{m}):
\begin{equation}
m_W = 80.3489 \pm 0.0133\; {\rm GeV} \;.
\end{equation}
The physical $W$ boson mass is 20.3 MeV less than the parameter $M_W$, which is nearly twice the uncertainty in the mass.

The top-quark mass and width are also extracted from experiment using the parameterization of Eq.~(\ref{Edepwidth}). The world-average values are $M_t = 172.57 \pm 0.29\; {\rm GeV}$ and ${\it \Gamma}_t = 1.42 + 0.19 - 0.15\; {\rm GeV}$. The width is sufficiently narrow that $M_t$ is equal to the physical top quark mass well within the uncertainties. In addition, the physical top quark mass is ambiguous by an amount of order $\Lambda_{QCD}\sim 200\; {\rm MeV}$ \cite{SW,H}

The Higgs boson width is expected to be so narrow ($\sim 4$ MeV) that the difference between the physical mass and the parameter $M_H$ is negligible.

\bigskip

\noindent {\bf 7. Quantum Field Theory Textbooks}

\bigskip

Although there are dozens of textbooks on quantum field theory, very few discuss unstable particles beyond calculating the decay rate of an unstable particle at leading order. These books play an important role in the way we pass knowledge from one generation to the next. Here I discuss several quantum field theory textbooks that have significant treatment of unstable particles.

The 1992 textbook of Lowell Brown \cite{Brown}, Sec.\ 6.3, gives a treatment of unstable particles that is very close to the modern treatment discussed in Sec.\ 2. However, he assumes $\Gamma \ll m$, both implicitly and explicitly, so he arrives at a relativistic Breit-Wigner formula that is valid only in this approximation, Eq.~(\ref{RIBW}) [see his Eq.~(6.3.24)]. The book does not include any discussion of the electroweak interaction, so we do not know if he would have treated the $Z$ boson any differently.

The 1995 textbook of Steven Weinberg \cite{Weinberg} includes Sec.\ 3.8 on ``Resonances.'' He begins by assuming the existence of a pole in the lower half of the complex energy plane, $\mu = E_R - \frac{i}{2}\Gamma$, and observes that this corresponds to a state of energy $E_R$ and decay rate $\Gamma$, as we showed in Sec.\ 2. He then shows that this leads to the Breit-Wigner resonance formula, Eq.~(\ref{BW}). If he had included an antiparticle pole in the upper half plane, he would have obtained the exact relativistic Breit-Wigner formula, Eq.~(\ref{resonance}). One of the examples he cites is the $Z$ boson, so he explicitly defines the $Z$ mass as the physical mass, $m_Z = E_R$.

The 1995 textbook of Peskin and Schroeder \cite{PS} has a discussion of unstable particles in Sec.\ 7.3. Without naming it, they define the mass in the ``on-shell'' scheme, Eq.~(\ref{OS}) [see their Eq.~(7.58)]. They arrive at the formula for a relativistic Breit-Wigner with an ``energy-dependent width,'' Eq.~(\ref{RBW}) [see their Eq.~(7.59)].\footnote{Following Ref.~\cite{BJ}, we derived Eq.~(\ref{BJpropagator}) by neglecting ${\rm Re}\,\Pi(p^2) - {\rm Re}\,\Pi(M^2)$. In Ref.~\cite{PS}, Eq.~(7.59) is obtained by expanding ${\rm Re}\,\Pi(p^2) - {\rm Re}\,\Pi(M^2) = (p^2-M^2){\rm Re}\,\Pi^\prime(M^2) + \cdots$, keeping only the first term, and defining $Z^{-1}=1+{\rm Re}\,\Pi^\prime(M^2)$.} They also arrive at a formula for the decay rate [their Eq.~(7.61)] which corresponds to the first term of the exact expression, Eq.~(\ref{mG}), and is thus accurate to NLO but not beyond.

The 2005 textbook of Michele Maggiore \cite{Maggiore} gives a heuristic derivation of the propagator of an unstable particle, with the pole at $\mu^2 = (m-\frac{i}{2}\Gamma)^2$. He then makes the approximation $\Gamma\ll m$ to arrive at the propagator that is valid in that approximation, Eq.~(\ref{approxprop}) [see his Eq.~(6.53)]. He then neglects the antiparticle pole to arrive at the Breit-Wigner resonance formula, Eq.~(\ref{BW}).

The 2007 textbook of Mark Srednicki \cite{Srednicki}, Ch.~25, defines the mass of an unstable particle as the ``on-shell'' mass, without calling it that. He arrives at the leading-order formula for the decay rate and states that this relation persists at higher orders in perturbation theory, which Eq.~(\ref{mG}) shows is not true. In both this book and Peskin and Schroeder, the confusion arises from the fact that the decay rate is determined by ${\rm Im}\,\Pi(\mu^2)$ [see Eq.~(\ref{width})], not ${\rm Im}\,\Pi(m^2)$.

The 2014 textbook of Matthew Schwartz \cite{Schwartz}, Sec.\ 24.1.4, also defines the mass of an unstable particle as the ``on-shell'' mass, Eq.~(\ref{OS}), although he refers to it as the ``real pole mass'' or the ``Breit-Wigner mass'' [see his Eq.~(24.49)]. He assumes $\Gamma \ll m$ throughout, so all equations should be understood as approximate. He arrives at a relativistic Breit-Wigner formula that is valid in this approximation, Eq.~(\ref{RIBW}) [see his Eq.~(24.51)]. At the end of the section he mentions the definition of the mass of an unstable particle via the complex pole in the propagator, Eq.~(\ref{pole}), and calls it the ``complex pole mass,'' but does not develop the idea.

Sidney Coleman's quantum field theory lectures from 1975-1976 are collected in a textbook published in 2019 \cite{Coleman}. Lecture 17 on unstable particles is an interesting mash-up of concepts. He defines the squared mass of an unstable particle as the zero of the real part of the denominator of the propagator as in Eq.~(\ref{mBW}) [see his Eq.~(17.7)], which in the mid 1970s is the ``Breit-Wigner mass,'' later called the ``on-shell'' mass.\footnote{He denotes the mass of the particle as $\mu$, which is unfortunately the symbol we are using for the complex pole position.} Working to NLO, he arrives at the propagator for a relativistic Breit-Wigner that is valid in the approximation $\Gamma \ll m$, Eq.~(\ref{RIBW}) [see his Eq.~(17.17)]. He then arbitrarily adds a NNLO term such that the pole in the propagator is located at $\mu^2 = (m-\frac{i}{2}\Gamma)^2$, without commenting on why he is doing this. Recall that just a few years earlier he had steered the hadronic physics community towards defining the mass and width this way. He then returns, without comment, to the relativistic Breit-Wigner that is valid in the approximation $\Gamma \ll m$, Eq.~(\ref{RIBW}) [see his Eq.~(17.18)], and shows that the mass is the oscillation frequency and the width is the decay rate. Rather than using contour integration as we have in Sec.~2, he approximates the Fourier transform using the method of stationary phase. That method tacitly assumes $\Gamma\ll m$,\footnote{The approximate size of the interval over which the phase of the $s$ integral [his Eq.~(17.42)] is stationary is $\sqrt{s_0/m}$, where $s_0$ is the proper time. One is therefore justified in approximating the function $e^{-\frac{1}{2}\Gamma s}$ at the stationary point $s_0$ if that function does not vary much over this interval, which implies $\Gamma\ll \sqrt{m/s_0}$. This, together with the explicit assumption $s_0m\gg 1$, implies $\Gamma \ll m$.} which is consistent with using Eq.~(\ref{RIBW}) as his starting point.

I hope that the next generation of quantum field theory textbooks will treat unstable particles in a fully modern way.

\bigskip

\noindent {\bf 8. Conclusions}

\bigskip

Peierls had the initial insight, in 1954, that an unstable particle is associated with a pole in the complex energy plane. This had to be rediscovered twice: in the early 1970s in the context of hadronic resonances, and in the early 1990s in the context of the $Z$ boson. We tend to think of knowledge as building up over time, but in this case it had to be rebuilt twice. It is also an example of a disconnect between two distinct communities within particle physics.

We now know that the description of unstable particles is incredibly simple. An unstable particle is associated with a pole at $\mu = m - \frac{i}{2}\Gamma$ in the complex energy plane, where $m$ is the mass and $\Gamma$ is the decay rate. In the idealized case that there is no nonresonant contribution to the amplitude, this produces a resonant cross section described by the celebrated Breit-Wigner formula [Eq.~(\ref{BW})],
\begin{equation}
|S|^2 \sim \frac{1}{(E-m)^2 + \Gamma^2/4}\;.
\end{equation}
If we desire a more accurate description, we should take into account that every particle is accompanied by an antiparticle, which is associated with a pole at $-\mu$. The idealized resonance formula is then given by Eq.~(\ref{resonance}),
\begin{eqnarray}
|S|^2 & \sim & \frac{1}{(p^2 - m^2 + \Gamma^2/4)^2+m^2\Gamma^2} \label{RBW2} \\
& = & \frac{1}{(E-m)^2 + \Gamma^2/4}\cdot\frac{1}{(E+m)^2 + \Gamma^2/4}
\end{eqnarray}
which is usually referred to as a relativistic Breit-Wigner formula, although the original Breit-Wigner formula is also valid at relativistic energies. In the final expression above we have gone to the rest frame of the particle and antiparticle to explicitly show the connection to the Breit-Wigner formula.
The relativistic Breit-Wigner formula, Eq.~(\ref{RBW2}), is almost always presented, either explicitly or implicitly, in the approximation $\Gamma\ll m$ [see Eq.~(\ref{RIBW})]. In the Breit-Wigner formulae above, $\Gamma$ is a constant; the concept of an ``energy-dependent width'' does not arise in our modern formulation of unstable particles.

In the real, nonidealized world there are also nonresonant contributions to amplitudes.  The pole position continues to define the mass and decay rate via $\mu = m - \frac{i}{2}\Gamma$.
All modern approaches to resonances (see Sec.\ 6 of Ref.~\cite{DD2}), including the complex mass scheme \cite{DDRW,DD1}, are based on the complex pole position, which is gauge-invariant and infrared safe \cite{GG}. Hopefully this knowledge will make it into the next generation of quantum field theory textbooks. The ``Breit-Wigner mass'' used in hadronic physics depends on the parameterization of the ``energy-dependent width,'' and the electroweak version, called the ``on-shell'' mass, is gauge dependent. The ``Breit-Wigner mass,''``on-shell'' mass, and ``energy-dependent width'' should all be regarded as historical curiosities and added to the list of discarded concepts in particle physics.

\newpage

\noindent {\bf Acknowledgements}

\bigskip

I am grateful for correspondence with Vernon Barger, Gordon Baym, Stan Brodsky, Bob Cahn, Randy Durand, Ayres Freitas, David Griffiths, Francis Halzen, Tao Han, Christoph Hanhart, Gordon Kane, Andreas Kronfeld, Peter Lepage, Heath O'Connell, Michael Peskin, Lee Pondrom, Chris Quigg, Jeff Richman, Jon Rosner, Ray Sawyer, and German Valencia.

\begin{figure}
\centering
\includegraphics[width=\textwidth]{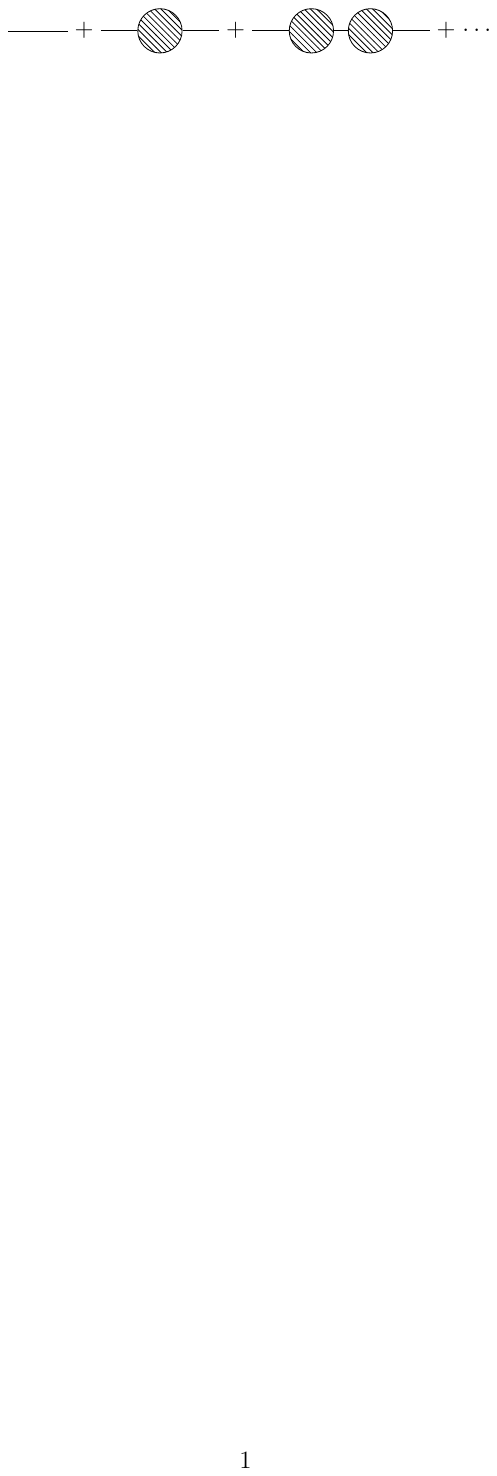}
\caption{Self-energy corrections to the propagator.}
\end{figure}

\begin{figure}
\centering
\includegraphics[width=\textwidth]{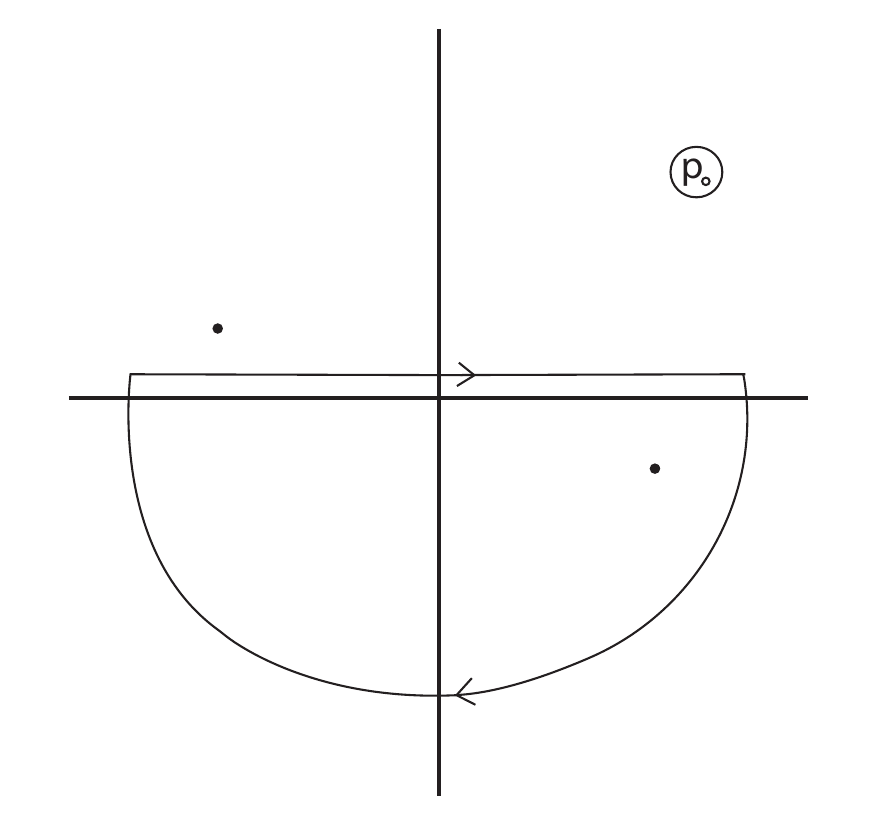}
\caption{Evaluating the Fourier transform of the propagator by closing the contour in the lower-half energy plane (for $t>0)$.}
\end{figure}

\end{document}